\documentclass{PoS}

\title{Exotic Discoveries in Familiar Places:\\ Theory of the Onia and Exotics}

\ShortTitle{Theory of the Onia and Exotics}

\author{\speaker{Richard F. Lebed}%
  \thanks{Work supported in part by the U.S.\ National Science Foundation under Grant No.\ PHY-1403891.}\\
  Arizona State University\\
  E-mail: \email{richard.lebed@asu.edu}}

\abstract{This brief overview presents an introduction to the
  heavy-quark exotic hadron candidates, $X$, $Y$, $Z$, $P_c$, believed
  to be tetraquark and pentaquark states.  After an abridged account
  of their discovery, we outline the various theoretical pictures
  employed to describe their structure, with special focus on the new
  {\em dynamical diquark\/} picture.}

\FullConference{16th International Conference on B-Physics at Frontier Machines\\
		 2--6 May 2016\\
		 Marseille, France}

\begin{document}

\section{Introduction}

Textbooks tell us that hadrons occur in two conventional types: $q\bar
q$ mesons and $qqq$ baryons.  Especially thorough texts also mention
the possibility of other, {\it exotic\/} color-neutral combinations:
$gg$ {\it glueballs}, $q\bar q g$ {\it hybrids}, $q\bar q q\bar q$
{\it tetraquarks}, $qqqq\bar q$ {\it pentaquarks}, and so on.  Yet
only in 2003 was the first particle clearly identified for which a
conventional assignment fails, the $X(3872)$~\cite{Choi:2003ue}.
Since then, almost 30 additional candidate exotics have been observed,
in both the tetraquark and pentaquark sectors.  In this brief
overview, we examine the major experimental discoveries, the evidence
for exoticness, and the variety of theoretical pictures developed to
describe their structure.

\section{The Crucial Importance of Quarkonium}

The quark model is now over half a century old.  Why did identifying
exotics take so long, and why did they first appear in the charmonium
($c\bar c$) sector, when so very many light hadron states are known?
In part, the answer is that the light sector (< 2.4~GeV) is rather
crowded.  It is easy for exotics made of $u,d,s$ to ``hide'' amongst
ordinary hadrons with the same $J^{PC}$ quantum numbers via mixing.
States $\pi_1(1400)$, $\pi_1(1600)$ with manifestly exotic $J^{PC} =
1^{-+}$ (hybrid meson candidates) have been observed, but must be
extracted from data through a complicated partial-wave analysis.
Occasionally, a strong signal for a truly exotic particle can arise,
even in multiple experiments, and later
disappear~\cite{Schumacher:2005wu}: In the early 2000's, evidence for
a candidate pentaquark $\Theta^+(1540)$ $K$-$N$ $s$-channel resonance
was eventually found to be primarily the result of unfortunate cuts
applied to the data, and likely also influenced by {\it kinematical
  reflections\/} from $t$-channel exchanges.  So when the true
breakthrough occurred in 2003, it was not instantly accepted by
everyone.

\begin{figure}[ht]
\includegraphics[angle=-90,origin=c,width=\textwidth]{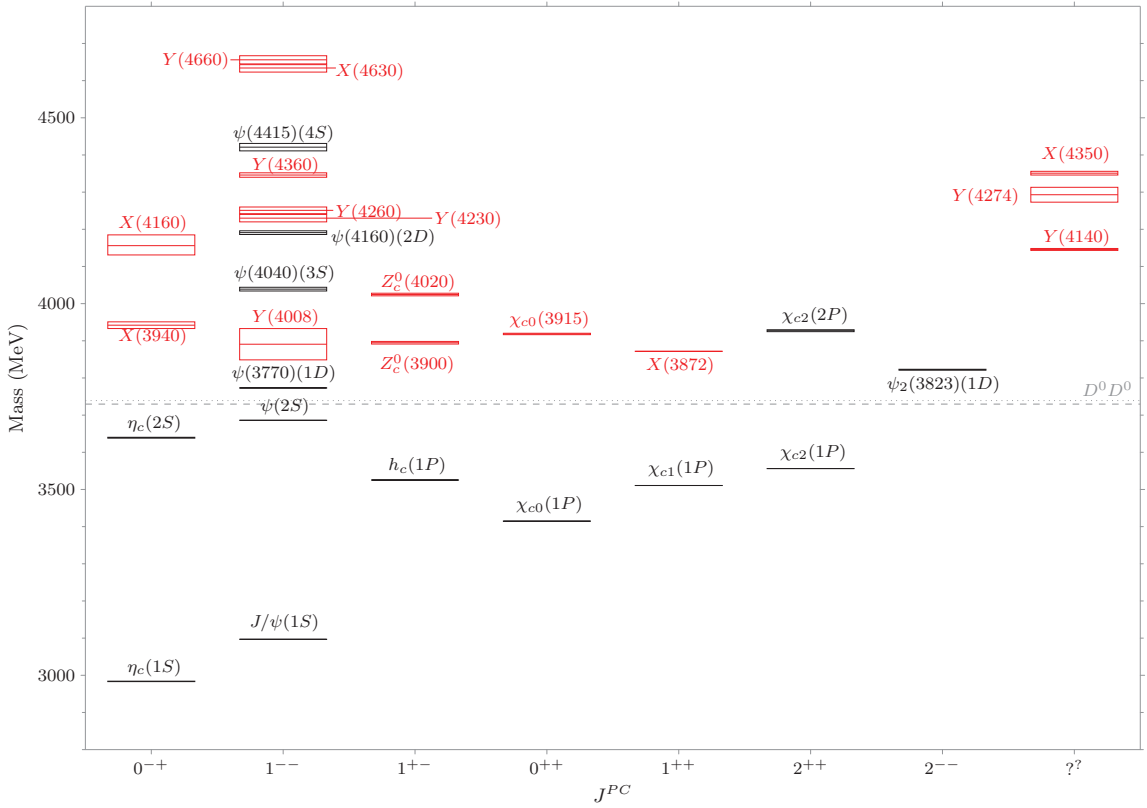}
\vspace{-12ex}
\caption{The neutral charmonium system.  Black lines indicate the
  conventional $c\bar c$ states predicted by quark-potential models
  that have been experimentally observed.  Red lines indicate the
  charmoniumlike exotic states.  Rectangles give the 1$\sigma$
  uncertainty of measured mass values.  The lowest open-charm ($D^0
  \bar D^0$) threshold is indicated as a dashed line.}
\label{Fig:ccNeutral}
\end{figure}

The key feature of the $c\bar c$ system (and by extension, systems
with $b$ quarks) leading to stronger statements about exotics
production is that the charm quark is the lightest one substantially
heavier than the typical QCD energy scale, $m_c \gg \Lambda_{\rm
  QCD}$.  As was immediately appreciated in the wake of the ``November
Revolution'' of 1974~\cite{Aubert:1974js,Augustin:1974xw} when the
first charmonium state $J/\psi$ was discovered, $m_c$ is large enough
to allow one to determine the spectrum of charmonium, and the QCD
running coupling at this scale, $\alpha_s (m_c) \simeq 0.3$, is small
enough to justify a simple quantum-mechanical treatment.  These facts,
even more than the discovery of scaling in deep-inelastic scattering a
few years prior, were instrumental in convincing the final skeptics of
the reality of quarks and color charges.  The most successful
treatment in this regard, the so-called ``Cornell
potential''~\cite{Eichten:1978tg,Eichten:1979ms}, models the strong
interaction between the $c\bar c$ pair as
\begin{equation}
V(r) = -\frac{k\alpha_s}{r} + br \, ,
\end{equation}
where the first term is the strong Coulomb-like one-gluon exchange,
and the second term represents the string or color flux tube-like
confining interaction.  Various small modifications ({\it e.g.},
spin-spin or spin-orbit terms) can be included, but once the
particular form of $V(r)$ is decided, it can be fed directly back into
the Schr\"{o}dinger equation to obtain the explicit $c\bar c$ spectrum
and wave functions.  Such models have successfully predicted the
entire $c\bar c$ spectrum below the lowest open-charm ($D^0 \bar D^0$)
threshold and also a number of states above it (black lines in
Fig.~\ref{Fig:ccNeutral}), and similarly in the $b\bar b$ system.
Possessing explicit wave functions also allows one to predict
radiative and hadronic decay rates, which generally agree well with
data.

The existence of many additional states in the region of neutral
$c\bar c$ states (red lines in Fig.~\ref{Fig:ccNeutral}) thus strongly
points to a large and complex system of {\it charmoniumlike\/}
exotics.  That they truly contain a $c\bar c$ valence pair is
evidenced by the fact that every observed decay channel contains a
$c\bar c$ pair.  Unless we have somehow grossly misunderstood
conventional $c\bar c$ systems, these extra states must certainly be
QCD exotics.  Moreover, as several {\em charged\/} charmoniumlike
states (discussed below) have been discovered, the neutral $c\bar c$
or $c\bar c g$ structures are insufficient to describe them all.

\section{Discovery of the Exotic Hadrons $X$, $Y$, $Z$, $P_c$}

The $X(3872)$ was discovered by Belle in 2003 as a resonance of $\pi^+
\pi^- J/\psi$ in the decay $B \to K (\pi^+ \pi^- J/\psi)$.  It is
worth recalling that the principal physics goal of Belle was not to
look for exotics but to find $CP$ violation in the neutral $B$ sector;
very likely, many more such surprise discoveries lie in store for the
$b$-physics community.  The existence of $X(3872)$ has since been
confirmed by multiple experiments.  Its quantum numbers are $J^{PC} =
1^{++}$, but the nearest conventional $c\bar c$ state, the
yet-undiscovered $\chi_{c1}(2P)$, is predicted to lie many 10's of MeV
higher.  Its measured mass is $m_{X(3872)} = 3871.69 \pm 0.17$~MeV,
astonishingly close to the $D^0 \bar D^{*0}$ threshold:
\begin{equation} \label{eq:thresh}
m_{X(3872)} - m_{D^{*0}} - m_{D^0} = -0.11 \pm 0.21 \, {\rm MeV} \, ,
\end{equation}
fueling endless speculation (first proposed
in~\cite{Tornqvist:2004qy}) that it is a $D^0\bar D^{*0}$ molecule.
In any case, it is believed to be a $c\bar c u\bar u$ state.  Its
width is remarkably small, $\Gamma_{X(3872)} < 1.2$~MeV, incompatible
with being an excited conventional $c\bar c$ state.  The label ``$X$''
emphasizes that its true structure and place in the family of hadrons
still remains unknown.

In 2005, the BaBar Collaboration discovered the second charmoniumlike
exotic, $Y(4260)$~\cite{Aubert:2005rm}.  They exploited the {\it
  initial-state radiation\/} (ISR) technique, in which the beam
$e^\pm$ pair emits a real photon before colliding; then the state
emerging from the collision vertex carries the $J^{PC} = 1^{--}$ of
the virtual photon.  $Y(4260)$ has been confirmed by other
experiments, and is charmoniumlike, as evidenced both by its mass and
decay modes, such as $\pi^+ \pi^- J/\psi$.  It is believed exotic
because all of the expected conventional $c\bar c$ states in the
$1^{--}$ channel up to the $4S$ $\psi(4415)$ have been seen (see
Fig.~\ref{Fig:ccNeutral}).  Subsequently, several other exotics have
been observed in the $1^{--}$ channel, all labeled $Y$.

The $Y(4260)$ was used in experiments by both the
BESIII~\cite{Ablikim:2013mio} and Belle~\cite{Liu:2013dau}
Collaborations for the first confirmation of a charged charmoniumlike
state, $Z_c^+ (3900)$, in 2013: It appears as a resonance in the decay
$Y(4260) \to \pi^- (\pi^+ J/\psi)$, and its quantum numbers have been
determined to be $J^P = 1^+$.  As a charged state, its minimal flavor
content is $c\bar c u\bar d$, meaning that it was the first manifestly
exotic state ever confirmed beyond $5\sigma$ by two experiments.
While news of this discovery should have become much more widely known
in the physics community, the cautionary example of the
$\Theta^+(1535)$ was still a recent memory; what if all of the
``states'' discovered up to this point turned out not really to be
states, but rather brilliant forgeries?

One important feature of a true resonance (in contrast to an
unfortunate statistical ``bump'' in cross-section data) is a
characteristic interference of its complex production amplitude $f(s)$
with nonresonant background.  In the idealized case of an isolated
Breit-Wigner resonance,
\begin{equation}
f(s) = \frac{\Gamma/2}{M - \sqrt{s} -i\Gamma/2} \, ,
\end{equation}
the scattering phase shift $\delta \equiv \arg f(s)$ increases rapidly
from 0 to $\pi$, creating a characteristic looping behavior in the
Argand plane.  In 2014, the LHCb Collaboration not only
confirmed~\cite{Aaij:2014jqa} the existence (at 13.9$\sigma$) of a
charged charmoniumlike state, $Z^-(4430) \to \pi^- \psi^\prime$ (first
seen by Belle in 2008~\cite{Choi:2007wga}) with $J^P = 1^+$ , but also
measured the phase shift in the resonant region and found it to be
compatible with the expected looping behavior.  While nature could be
especially unkind and generate this behavior even in the absence of a
resonance, such an accident seems unlikely, and moreover, further
tests by LHCb~\cite{Aaij:2015zxa} support that $Z^-(4430)$ is not
merely a kinematical reflection.  As of 2014, one could state with
some confidence that we had entered the Age of the Third Hadron.

Indeed, one may also say that the Age of the Fourth Hadron followed
only one year later.  In 2015, LHCb observed~\cite{Aaij:2015tga} the
first two baryonic charmoniumlike resonances, $P_c^+ (4380)$ (at
9$\sigma$) and $P_c^+ (4450)$ (at 12$\sigma$) decaying into $J/\psi \,
p$, and therefore possessing pentaquark $c\bar c uud$ quantum numbers.
While the $J^P$ assignments are not yet completely established, the
preferred solution gives opposite parities and one state with spin 3/2
and the other with spin 5/2.  Again, rapid phase variation in the
vicinity of the peaks, consistent with resonant behavior, was observed
[although resembling the ideal Breit-Wigner behavior rather less for
the $P_c^+(4380)$].

This discussion has been, of necessity, extremely brief and does not
mention all the other exotic candidates, the supporting experiments,
the conflicting signals, and the dead ends.  Nor does it mention that
a few exotic candidates have been observed in the $b\bar b$ sector as
well.  An expansive discussion appears in the recent review
Ref.~\cite{Chen:2016qju}.  As of the time of the BEAUTY 2016
Conference, 29 exotics had been observed, of which 24 are in the
$c\bar c$ sector, 4 are in the (much less explored) $b\bar b$ sector
(See, {\it e.g.},~\cite{Belle:2011aa,Krokovny:2013mgx}), and 1 in the
$bsud$ sector~\cite{D0:2016mwd}.  Of these, 12 have been confirmed by
a second experiment, and only 1 of the other 17 (the last) has been
challenged by strong contrary evidence~\cite{LHCb:2016ppf}.  Of
particular interest to this conference, a large number of
$b$-containing exotics almost certainly await discovery.

\section{How Are the Exotics Assembled?}

The existence of tetraquark and pentaquark states having been
established experimentally, we briefly explore their structure as
composite multiquark particles.  On the order of 1000 theoretical
papers have appeared since the discovery of the $X(3872)$, so it will
be impossible to include any but the earliest references to a
particular approach, and for specificity, we focus on the tetraquarks.

Given a $q \bar q q \bar q$ system, simple SU(3) color algebra shows
that color singlets can be formed in exactly two ways, which
correspond to the two possible independent $q\bar q$ pairings.  While
it is tempting to describe these pairings as corresponding to the
formation of hadronic molecules [{\it e.g.}, in the case of $X(3872)$,
either $D^0 \bar D^{*0}$ or $J/\psi \, \omega$], group theory alone
does not determine hadronic structure or dynamics.  The quarks can
segregate themselves spatially in several other ways that do not
resemble a molecular configuration.  Let us first discuss the popular
non-molecular pictures.

Charmoniumlike {\it hybrids\/} were first modeled in
Ref.~\cite{Giles:1976hh}).  They may explain some of the neutral
exotics, but not the $Z^\pm$.  Furthermore, for most values of
$J^{PC}$, hybrids are expected from lattice QCD to lie higher in mass
than the observed exotics~\cite{Braaten:2014qka}.

The {\it hadrocharmonium\/} picture~\cite{Voloshin:2007dx} assumes
that the heavy $c\bar c$ pair in a tetraquark resides at the center of
a cloud provided by the lighter two quarks.  Such models are best at
describing decays of the exotic into conventional charmonium, but are
somewhat less successful for open-charm decays.

The {\it diquark\/} picture (sometimes confusingly referred to as
``tetraquark'') is based on the idea that quarks have a
little-appreciated attractive channel: In addition to the well-known
color-singlet $q\bar q$ combination, a $qq$ pair at short distance
exhibits a strong attraction, fully one-half as large, into a
color-$\bar{\bf 3}$.  One can imagine multiquark hadrons with compact
diquark ($qq$) or antidiquark ($\bar q \bar q$) components.  The
spectroscopy of such states has been developed in
Refs.~\cite{Maiani:2004vq,Maiani:2014aja} and, for example, gives a
simple explanation for the $Y$ states as $L=1$ excitations of the
ground states such as $X(3872)$.  However, diquark models also tend to
greatly overpredict the number of states expected.

It is even possible for apparent resonances not to be true states at
all, but rather an effect caused by the rapid opening of a nearby
two-meson threshold.  The complex analyticity properties of the
associated two-point function create a effective attraction that can,
in certain circumstances, produce a cusp-shaped profile resembling in
all respects a resonance.  The most well-known of these {\it kinematic
  effects\/} is the {\it cusp effect\/} first suggested for the heavy
exotics in Ref.~\cite{Bugg:2004rk}; however, {\it rescattering
  effects\/} due to the production of loops of virtual particles can
have similar effects.  Nevertheless, not every exotic lies especially
close to a threshold [the $Z^-(4430)$ is a prime example], and states
as narrow as the $X(3872)$ are very difficult to explain purely as a
kinematical effect.

We return at last to the simple {\it hadronic molecule\/} picture,
first proposed for $c\bar c$ systems
in~\cite{Voloshin:1976ap,DeRujula:1976zlg}.  A number of the exotics
lie suspiciously close to hadron thresholds [{\it e.g.},
Eq.~(\ref{eq:thresh})].  However, not all are, as has just been noted
for the $Z^-(4430)$.  Moreover, some $XYZ$ states lie slightly {\em
above\/} a hadronic threshold [{\it e.g.}, $Y(4260)$ lies about 30~MeV
above the $D_s^* \bar D_s^*$ threshold], creating the awkward problem
of a bound state with positive binding energy.  In the well-studied
case of the $X(3872)$, one would expect a state with such a tiny
binding energy---the central value is 20 times smaller than the
binding energy of a deuteron, already considered weakly bound---to be
especially delicate.  The $X(3872)$ might appear in $B$ decays, but
one would expect them to be blown apart at the high energies of
collider experiments.  Plenty of $D^0$ and $\bar D^{*0}$ mesons should
be produced in {\it prompt production\/} in the central interaction
region, but very few should be produced close enough in phase space to
coalesce into a molecular configuration, even including strong
final-state interaction effects.  And yet, CDF~\cite{Abulencia:2006ma}
and CMS~\cite{Chatrchyan:2013cld} produced $X(3872)$ as copiously as
conventional $c\bar c$ states.  Charmoniumlike hadronic molecules may
certainly exist, but the $X(3872)$ does not seem to fit the expected
profile.

It is entirely possible that no single structure accommodates all of
the exotic states.  Some could be molecules, some hybrids, some
kinematical effects, {\it etc.}  Or they could be quantum-mechanical
mixtures of several of these.  For example, allowing both molecular
and conventional $c\bar c$ components for the $X(3872)$ is very
helpful in accommodating its currently known properties.  Each picture
deserves continued study to map out its successes, shortcomings, and
limits of applicability.

\section{A New Dynamical Picture for the $X$, $Y$, $Z$, $P_c$}

A feature shared by all of the aforementioned pictures is a reliance,
either explicitly or implicitly, on essentially static structures
({\it e.g.}, the formation of recognizable mesonic degrees of freedom,
or diquarks in well-defined orbitals).  They do not obviously
incorporate the full complexity of QCD dynamics in what are, after
all, rather short-lived states.  This author and collaborators
therefore recently developed a new picture~\cite{Brodsky:2014xia} to
accommodate these facts.

The claim is that at least some of the observed tetraquark states are
bound states of diquark-antidiquark pairs, as suggested above.
However, unlike other familiar bound states, the components are not in
a static molecular configuration.  Rather, the diquarks are created
with a large relative momentum, and rapidly separate from each other
(hence the name {\it dynamical diquark\/} picture).  Since the
diquarks are not color neutral, they cannot, by confinement, separate
arbitrarily far.  Their kinetic energy is converted into the potential
energy of the color flux tube connecting them, until the pair is
brought to rest at some substantial separation, at which point
hadronization occurs though the large-distance tails of meson wave
functions stretching from the $q$'s of the diquarks to the $\bar q$'s
of the antidiquarks.  But these tails are exponentially suppressed,
leading to suppressed decay transitions and hence, observably narrow
widths.  Why should this type of hadronization occur, rather than a
simple breaking of the flux tube at this point, creating an additional
$q\bar q$ and resolving the system via baryon-antibaryon creation?
The lowest such threshold is $\Lambda_c \bar \Lambda_c$ at 4573~MeV,
and indeed, the lowest exotic candidate [$X(4630)$] above this
threshold is observed~\cite{Pakhlova:2008vn} to decay only into
$\Lambda_c \bar \Lambda_c$.

As an example, consider the production of $Z^+(4430)$ in nonleptonic
$\bar B^0 = b \bar d$ decay (production in $b$-factories and colliders
is similar).  The weak transition $b \to c s \bar c$ suggests the
obvious ``color-allowed'' open-charm decay route $(b \bar d) \to (c
\bar d) (s \bar c)$ = $D^{(*)+} \bar D_s^{(*)-}$, where parentheses
indicate the compact quark combinations.  This particular quark
transition occurs about 22\% of the time, and the given 2-body decays
amount to about 5\% of $\bar B^0$ decays~\cite{Agashe:2014kda}.  In
such decays, the c.m.\ momentum of the hadrons is $\simeq$ 1700~MeV,
simply reflecting the large energy release due to $m_b \gg m_c$.

Of course, several other decay mechanisms are available, including
color-suppressed decays, penguins, and so on.  In the dynamical
diquark picture, in some fraction of decays the formation of a
diquark-antidiquark pair is favorable.  In the case of $\bar B^0$, let
the weak decay $b \to c s \bar c$ be accompanied by the creation of an
additional $u\bar u$ pair from the vacuum.  Some fraction of the time,
the $(uc)$ and $(\bar c \bar d)$ pairs appear in sufficient proximity
to form color-triplet diquarks rather than color-singlet $q\bar q$
pairs.  The remaining $(s\bar u)$ forms a $K^-$, so the production
process, just as is experimentally observed, is $\bar B^0
\to K^- Z^+(4430)$; were the diquarks color neutral, this process
would be a conventional 3-body hadronic decay.  Instead, the $(uc)$
and $(\bar c \bar d)$ diquarks separate until forced to come to rest
and hadronize as described above, into the observed pair $\psi^\prime
\pi^+$; this configuration defines the $Z^+(4430)$.

To support this picture, note a remarkable experimental feature of the
$Z^+(4430)$: It heavily prefers to decay to the $J^{PC} = 1^{--}$ $2S$
state $\psi^\prime$ rather than the $1S$ state $J/\psi$ by at least a
factor of 10~\cite{Chilikin:2014bkk}, despite the significantly larger
phase space for the latter.  This fact is easy to understand if one
calculates the final separation $r_Z$ of the diquark-antidiquark pair
using, say, the Cornell potential; one finds $r_Z = 1.16$~fm.  In
comparison, the same potential gives charge radii of the $c\bar c$
states, $\left< r_{J/\psi} \right> = 0.39$~fm and $\left<
  r_{\psi^\prime} \right> = 0.80$~fm.  A simple matter of greater wave
function overlap for the $\psi^\prime$ in the dynamical diquark model
explains its preferred status in $Z^+(4430)$ decay.

Finally, a few words about the $P_c^+$ states: The color attraction
for ${\bf 3} \times {\bf 3} \to \bar {\bf 3}$ applies not just to
diquarks, but can be compounded successively to triquarks and beyond.
In Ref.~\cite{Lebed:2015tna}, the discovery channel $\Lambda_b \to
P_c^+ K^-$ is seen to be an exact analogue to the $\bar B^0 \to
Z^+(4430) K^-$, except that the diquark $(\bar c \bar d)$ is replaced
by the color-{\bf 3} {\it antitriquark\/} $[\bar c (ud)]$.

\section{The Present and the Future}

The past two years have provided confirmation of the existence of
tetraquarks and observation, at high statistical significance, of the
pentaquark, the third and fourth classes of hadron.  To date, almost
30 such states ($X,Y,Z,P_c$) have been observed.  Several theoretical
pictures have been developed to describe their structure (molecules,
hybrids, diquark-antidiquark states, hadrocharmonium, and kinematic
effects), and while each one has appealing features, none of them
appears to universally explain all of the new states.  We have also
described the new ``dynamical diquark'' picture, designed to allow
creation of a spatially large hadronic system still strongly bound by
color forces.

The field of heavy-quark QCD exotics is nonetheless strongly
data-driven.  $b$-physics experiments will continue to lead the way in
the development of this rich subject (as evidenced by reports in this
conference on the latest exotics results from ATLAS, BESIII, CMS, and
LHCb~\cite{LatestExpts}, and prospects for the BelleII
upgrade~\cite{BelleUpgrade}), and a final understanding of these
states will become clear.

\end{document}